\documentclass[sigconf,nonacm]{acmart}

\usepackage{amsthm}
\usepackage{array,multirow}
\usepackage{booktabs} 
\usepackage{xcolor}
\usepackage{subcaption}
\usepackage{caption}
\usepackage{xspace}
\usepackage{graphicx}
\usepackage{multirow}
\usepackage{siunitx}
\usepackage{adjustbox}
\usepackage{hyphenat}
\usepackage{makecell}
\usepackage{colortbl}
\usepackage{fancyhdr,graphicx,amsmath}
\usepackage[ruled,vlined]{algorithm2e}
\usepackage[referable]{threeparttablex}
\usepackage{listings}
\usepackage{listingsutf8}
\usepackage[normalem]{ulem}

\SetCommentSty{mycommfont}

\newcommand{\systemnamenf}{M2}
\newcommand{\systemname}{{\em M2 }}
\newcommand{\systemnamens}{{\em M2}}

\newcommand{\mmjoin}{{\em MSHJ }}
\newcommand{\mmjoinns}{{\em MSHJ}}

\newcommand{\eg}{{\em e.g.}}
\newcommand{\reminder}[1]{{\color{red} $\rightarrow$ #1}}

\def\code#1{\texttt{#1}}

\newcommand{\keyword}[1]{{\textit{\textbf{#1.}}}}

\newtheoremstyle{mystyle}
  {}
  {}
  {\itshape}
  {}
  {\bfseries}
  {.}
  { }
  {}
\theoremstyle{mystyle}

\newtheorem{definition}{Definition}

\newtheorem{example}{Example}
\newenvironment{qedexample}
{\begin{example}}    
{\qed               
\end{example}}      

\AtBeginDocument{%
  }

\setcopyright{acmlicensed}
\copyrightyear{2018}
\acmYear{2018}
\acmDOI{XXXXXXX.XXXXXXX}

\acmConference[CIDR'26]{Make sure to enter the correct
  conference title from your rights confirmation email}{January 18--21,
  2026}{Chaminade, USA}

\acmISBN{978-1-4503-XXXX-X/2018/06}

\setcopyright{none} 
\renewcommand\footnotetextcopyrightpermission[1]{} 

\begin{document}

\title{\systemnamenf: An Analytic System with Specialized Storage Engines for Multi-Model Workloads}

\def\snudept{Computer Sci \& Eng}
\def\snu{Seoul National University}
\def\snuaddress{1 Gwanak-ro, Gwanak-gu}
\def\snucity{Seoul}
\def\snucountry{Republic of Korea}
\def\snuaffl{
  \institution{\snu}
  \city{\snucity} 
  \country{\snucountry}
}

\author{Kyoseung Koo}
\affiliation{\snuaffl}
\email{koo@dbs.snu.ac.kr}


\author{Bogyeong Kim}
\affiliation{\snuaffl}
\authornote{The author contributed to this work when he was affiliated with Seoul National University.}
\email{bgkim@dbs.snu.ac.kr}

\author{Bongki Moon}
\affiliation{\snuaffl}
\email{bkmoon@snu.ac.kr}

\renewcommand{\shortauthors}{Kyoseung Koo, Bogyeong Kim, and Bongki Moon}

\begin{abstract}

Modern data analytic workloads increasingly require handling multiple data models simultaneously.
Two primary approaches meet this need: polyglot persistence and multi-model database systems.
Polyglot persistence employs a coordinator program to manage several independent database systems but suffers from high communication costs due to its physically disaggregated architecture.
Meanwhile, existing multi-model database systems rely on a single storage engine optimized for a specific data model, resulting in inefficient processing across diverse data models.
To address these limitations, we present \systemnamens, a multi-model analytic system with integrated storage engines.
\systemname treats all data models as first-class entities, composing query plans that incorporate operations across models.
To effectively combine data from different models, the system introduces a specialized inter-model join algorithm called {\em multi-stage hash join}.
Our evaluation demonstrates that \systemname outperforms existing approaches by up to $188 \times$ speedup on multi-model analytics, confirming the effectiveness of our proposed techniques.

\end{abstract}

%
%


\maketitle

\section{Introduction}

Unlike traditional workloads based on a single data model, modern data analytic workloads increasingly require the ability to handle multiple data models simultaneously.
In the fields of machine learning and artificial intelligence, in particular, workloads typically involve processing array data for matrix computations or linear algebra algorithms together with structured and semi-structured data.
Consider, for example, an e-commerce recommendation system that builds a non-negative matrix factorization model. 
In this scenario, order and review data, both stored in a document-oriented model, are combined to construct a recommendation matrix that represents rating values for products by customers.
This matrix is then fed into the factorization algorithm to generate recommendations.
To extract insights such as identifying products that a customer would like, the reconstructed matrix is converted into a relational format for further analysis using relation operations.
This data analysis pipeline requires handling three distinct data models, namely, relational, document-oriented, and array models.
There are many use cases reported in the literature for combining an array data model with structured and semi-structured data models~\cite{recommendersystem, FERDOUSI201754, xian2020urban}.

One way of processing such a multi-model analytic workload is to rely on a polyglot persistence system.
Polyglot persistence enables us to adopt multiple database systems simultaneously for the processing of complex workloads~\cite{polyglot2022vldb}.
In this approach, each database system is responsible for processing data in its own data model, while a coordinator module typically orchestrates query execution across them.
Unfortunately, however, the coordinator and the underlying database systems are physically disaggregated. 
Consequently, a substantial amount of communication overhead is incurred inevitably to transfer intermediate and final result data among the coordinator and the database systems~\cite{dziedzic2016data,kim2022m2bench}.
The overhead becomes aggravated as the volume of data involved in the workload increases.

Another approach has emerged that focuses on handling multiple data models within a single database system, which can be referred to as a single-engine multi-model database system~\cite{lu2019multi}.
Such a multi-model database system employs a single storage engine dedicated to a particular data model and attempts to accommodate all the supported data models with the storage engine.
AgensGraph, for example, utilizes PostgreSQL to support its query processing for data not only in the relational model but also in the graph model.
Evidently, this single storage engine approach suffers from non-trivial performance degradation due to the lack of direct support from the most relevant data model to each individual operation in a query plan.
Several works reported that using non-native storage engines to process a data model shows inferior execution time compared to those with native storage engines~\cite{kim2022m2bench,thomas_comparative_2018,koo2024prevision}.

This paper presents a prototype of a new multi-model database system called \systemname that can plan and execute a multi-model query with multiple underlying storage engines.
\systemname addresses the limitations intrinsic to the polyglot persistence and single-engine multi-model database systems by tightly integrating multiple storage engines together.
If a query involves operations defined in two or more distinct data models, then the query plan is segmented into partitions by models.
Each query partition is then processed efficiently by a storage engine dedicated to and optimized for the corresponding data model.
The current implementation of \systemname supports query processing across the relational, document-oriented, and array data models but can be extended to include additional data models such as a graph model.
\systemname can be considered a true {\em multi-engine multi-model} database system as it treats every supported data model as a first-class entity with equal priority.

\begin{figure}[t!]
    \centering
    \includegraphics[width=0.85\linewidth]{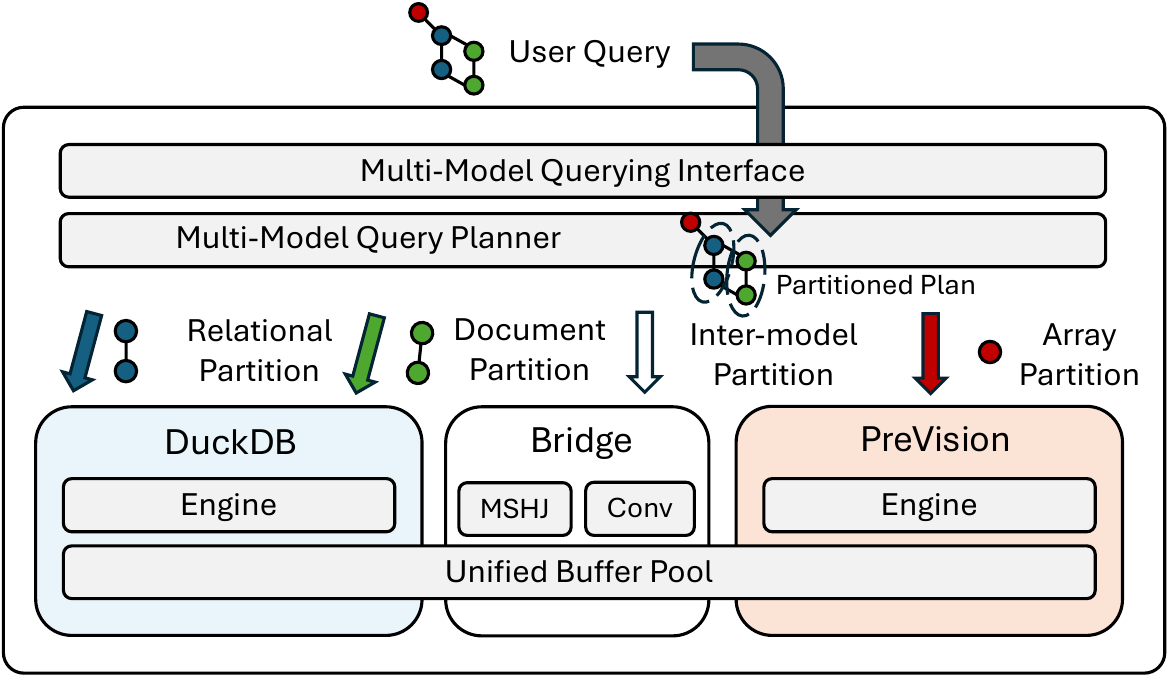}
    \caption{Architectural overview of \systemname}
    \Description{Architectural overview of the system. \reminder{storages and bridge are important}}
    \label{fig:overview}
\end{figure}

Figure~\ref{fig:overview} depicts the system architecture of \systemnamens.
The system runs on two storage engines: an augmented version of DuckDB~\cite{duckdb} for relational and document-oriented data and PreVision~\cite{koo2024prevision} for array data.
DuckDB is an open-source relational database system, and we have augmented it with additional functionalities for document-oriented data processing and interoperability with \systemnamens.
PreVision offers efficient multi-dimensional array processing capabilities with I/O optimized storage management.
These two storage engines are tightly integrated, and they \textit{share a common buffer pool} to improve the utilization of buffer space.
Furthermore, a dedicated bridge module is added to facilitate the cooperation of the two storage engines.
This bridge module is responsible for data conversion and transmission between them for inter-model operations such as {\em multi-stage hash join} that will be described later.

%
%

The remainder of this paper is organized as follows.
Section~\ref{sec:model_op} describes the data models and operations currently supported by the system.
Section~\ref{sec:storage} discusses the query execution flow and the unified buffer pool.
Section~\ref{sec:join} presents the multi-stage hash join algorithm.
An experimental evaluation is demonstrated in Section~\ref{sec:eval}, and related literature is summarized in Section~\ref{sec:related_work}.
Lastly, Section~\ref{sec:conclusion} concludes this work.

\lstdefinelanguage{json}{
    basicstyle=\ttfamily,
    showstringspaces=false,
    breaklines=true,
    frame=single,
    literate=
     *{\{\}}{\textbraceleft\textbraceright}1
      {\[\]}{\textbracketleft\textbracketright}1
      {,}{,}1
      {:}{:}1
      {\"}{\textquotedbl}1
}

\lstdefinestyle{jsonstyle}{language=JSON,
  basicstyle={\footnotesize\ttfamily},
  belowskip=3mm,
  breakatwhitespace=true,
  breaklines=true,
  classoffset=0,
  columns=flexible,
  commentstyle=\color{dkgreen},
  framexleftmargin=0.25em,
  keywordstyle=\color{teal},
  numbers=none, 
  numberstyle=\tiny\color{gray},
  showstringspaces=false,
  stringstyle=\color{darkgray},
}

\lstdefinestyle{pseudocode}{language=JSON,
  basicstyle={\footnotesize\ttfamily},
  belowskip=3mm,
  breakatwhitespace=true,
  breaklines=true,
  classoffset=0,
  columns=flexible,
  commentstyle=\color{dkgreen},
  framexleftmargin=0.25em,
  keywordstyle=\color{teal},
  numbers=left,
  numberstyle=\tiny\color{gray},
  showstringspaces=false,
  stringstyle=\color{darkgray},
}

\section{Data Model and Operation}
\label{sec:model_op}

This section outlines the supported data models (Section~\ref{subsec:model}) and the operations available for each data model, as well as the inter-model operations that work across different data types (Section~\ref{subsec:operation}).

\subsection{Model Definition}
\label{subsec:model}

We formally define the data models used in \systemnamens, beginning with their fundamental elements (record, document, and cell) and extending to their respective sets (relation, collection, and array).

\begin{definition}[\textbf{Record}]
A $record\ r$ is \textit{a list of attribute values} with a fixed number of attributes.
That is, $r = \{v_1, v_2, \ldots, v_n\}$, where $v_i$ is the $i$-th attribute value and $n$ is fixed.
\end{definition}

\begin{definition}[\textbf{Relation}]
A $relation\ R$ is \textit{a set of records} where all records share the same number of attributes and consistent attribute types.
That is, $R = \{r_1, r_2, \ldots, r_m\}$, where $\forall r_j \in R$, the number of attributes $n$ is fixed, and the type of each attribute remains consistent.
\end{definition}

A record represents a row in a relational database table, and a relation corresponds to a table containing multiple rows with the same schema.
A value type can be a list: it is often called an array, but this paper distinguishes those as the array model is defined later.

\begin{definition}[\textbf{Document}]
A $document\ d$ is \textit{a single-attribute record} whose value stores a list of key-value pairs.
That is, $d = \{(a_1, v_1),$ $(a_2, v_2),$ $\ldots,$ $(a_n, v_n)\}$, where $a_i$ is a key, $v_i$ is the corresponding value, and $n$ is the number of key-value pairs.
Within a document, any attribute value may itself contain another document, allowing for nested structures.
\end{definition}

\begin{definition}[\textbf{Collection}]
A $collection\ C$ is a set of documents, defined as $C = \{d_1, d_2, \ldots, d_m\}$.
\end{definition}

Documents represent JSON objects in document databases, where each object contains flexible key-value pairs.
For instance, the document below includes two key-value pairs.
The second pair includes a ``geometry'' field with a nested document. 
Other documents in the same collection may or may not contain identical fields.

\begin{lstlisting}[style=jsonstyle]
    {"id": 1015, "geometry":
        {"type": "Point","coordinates": [126.952511, 37.449527]}}
\end{lstlisting}

\begin{definition}[\textbf{Cell}]
A $cell\ c$ is \textit{a record} in which a subset of the attributes represents dimensions.
That is, $c = \{a'_1,$ $a'_2,$ $\ldots,$ $a'_d,$ $a_1,$ $a_2,$ $\ldots,$ $a_n\}$, where $a'_i$ is the $i$-th dimension value, $d$ is the number of dimensions, $a_i$ is the $i$-th non-dimension attribute value, and $n$ is the number of non-dimension attributes.
All dimension values are represented as unsigned integers.
\end{definition}

\begin{definition}[\textbf{Array}] 
An array $A$ is a set of cells arranged in a $d$-dimensional structure, defined as  $A = \{c_1, c_2, \ldots, c_m\}$.
The size of $A$ is defined as $AS = (l_1, l_2, \ldots, l_d)$, where each $l_i = MAX(a'_i \in c), c \in A$ represents the length of the $i$-th dimension.
Each cell $c \in A$ is uniquely identified by its dimension attributes.
\end{definition}

Cells are records containing both values and their dimensional coordinates. 
These coordinates may be conceptual rather than explicitly stored in the physical layout. 
For instance, in \systemnamens's array storage engine, PreVision, cell coordinates in dense arrays are derived from the cell's spatial location rather than stored explicitly.
Arrays are multi-dimensional data such as vectors, matrices, or raster data used in machine learning and scientific computing.
Each array maintains explicit size information for each dimension, as this is essential for determining the output shapes of array operations.

\subsection{Interface and Operation}
\label{subsec:operation}

Listing~\ref{lst:pseudocode} demonstrates how \systemname can be used to predict the top 10 favorable products for a given customer during targeted marketing campaigns.
The query begins by accessing \code{order} and \code{review} collections which are stored in a document-oriented model (Line~\ref{alg:ex:col_open}). 
A document in the order collection includes a customer identifier (\code{cid}), while a review document contains a product identifier (\code{pid}) and a rating value.
To obtain rating values for each customer and product, these collections are joined on order identifiers (\code{oid}) using the \code{join} function, followed by the \code{project} function to filter only relevant dimensions and rating values (Line~\ref{alg:ex:col_proc}).
Before being fed into the non-negative matrix factorization algorithm, the set of rating values is converted into a matrix by the \code{toArray} function with names for dimensions and a value (Line~\ref{alg:ex:toarr}). 
By using the number of tuples in customer and product tables and the \code{rand} function, the factor matrices \code{W} and \code{H} are initialized (Lines~\ref{alg:ex:nmf_init_start}-\ref{alg:ex:nmf_init_end}), and the algorithm updates these matrices (Lines~\ref{alg:ex:nmf_run_start}-\ref{alg:ex:nmf_run_end}). 
To focus the analysis on specific customer-product pairs of interest (for targeted marketing campaigns), the \code{interest} table containing customer and product identifiers is opened (Line~\ref{alg:ex:final_tblopen}).
Once the recommendation matrix is reconstructed (Line~\ref{alg:ex:final_recon}), the \code{join} function takes \code{interest} which is in a relational model and \code{filled} in an array model, producing equi-joined relational data as an output (this inter-model join operation will be discussed later).
Among the joined rating values, the ones with the cid = 3 are selected, and the top ten items are extracted (Line~\ref{alg:ex:imj}). 
Finally, the \code{execute} function is called to get results from the described query above (Line~\ref{alg:ex:exec}).

\begin{lstlisting}[
  float,
  floatplacement=t,
  style=pseudocode,
  escapechar=|,
  caption={Example pseudocode for illustration},
  label={lst:pseudocode},
  belowskip=-2em]
order, review = openCollection('order'), openCollection('review')   |\label{alg:ex:col_open}|
ratings = review.join(order, 'review.oid = order.oid').project('cid, pid, rating')   |\label{alg:ex:col_proc}|
X = ratings.toArray({'cid', 'pid'}, {'rating'})   |\label{alg:ex:toarr}|

rank, num_iter = 10, 3   |\label{alg:ex:nmf_init_start}|
customer_cnt, product_cnt = openTable('customer').count(), openTable('product').count()
W, H = rand({customer_cnt, rank}), rand({rank, product_cnt})   |\label{alg:ex:nmf_init_end}|
for i in range(num_iter):   |\label{alg:ex:nmf_run_start}|
  W = W * ((X @ H.T) / (W @ H @ H.T))
  H = H * ((W.T @ X) / (W.T @ W @ H))   |\label{alg:ex:nmf_run_end}|

interest = openTable('interest') |\label{alg:ex:final_tblopen}|
filled = W @ H |\label{alg:ex:final_recon}|
result = filled.join(interest, 'interest.cid = filled.cid AND interest.pid = filled.pid', RELATIONAL).filter('cid = 3').sort('rating DESC').limit(10) |\label{alg:ex:imj}|

execute(result)  |\label{alg:ex:exec}|
\end{lstlisting}

As depicted in the example, an \systemname query consists of a series of operations, including model-specific operations and inter-model operations (Table~\ref{table:operations} displays supported operations).
Model-specific operations take a specific model as input and produce the same model as output.
For the relational model, common relational operations, which can be expressed using Structured Query Language (SQL), are provided.
For the document model, in addition to the fundamental relational operations, \systemname provides the dot processing functionality for accessing values in nested documents and supports the unwind operation to unnest lists, which are commonly used in semi-structured data analysis.
The array model supports operations for linear algebra and raster processing.

In contrast to the model-specific operations, inter-model operations process data across model boundaries.
Model conversion operations transform data from one model into a target model, and  inter-model joins combine data from different models for integrated analysis. 
To precisely specify the semantics of the inter-model join in \systemnamens, we provide a formal definition.

\begin{definition}[\textbf{Inter-model join}] 
An inter-model join is a binary operation that combines two \textbf{records} from \textbf{different models} when they satisfy a given condition, producing results in one of the input models.
That is, $R \bowtie S = \{ f(r \cup s) \mid r \in R \wedge s \in S \wedge g(r, s) \wedge m(R) \neq m(S)\} $, where $R$ and $S$ are sets of records, $f$ is a projection function to produce the output model, $g$ is a join condition function, and $m$ maps the given argument to its associated model.
\end{definition}

For example, in our earlier e-commerce analysis, we joined a table with an array using the \code{join} function, matching tuples and cells based on equal identifiers.
Inter-model joins differ from traditional relational joins in two key aspects. 
First, they operate on heterogeneous data models rather than a single model, requiring distinct access methods for each model type.
Second, the output can be one of the input models. 
Note that the cross-model join supported by Polypheny-DB~\cite{polypheny} can be seen as an inter-model join with the fixed projection function producing a relational model.

Currently, \systemname utilizes two methods for inter-model joins: a \textit{multi-stage hash join} for array-involving equi-joins and a method relying on a relational join operation for the rest.
We will discuss the multi-stage hash join approach in Section~\ref{sec:join}.

\begin{table}[t]
	\small
    \centering
    \caption{Provided operations}
    \begin{tabular}{c|c}
        \toprule 
        \textbf{Model} & \textbf{Operation} \\ 
        \midrule
        Relational & \makecell[c]{Filter, Project, Sort, Limit, Aggregate, Union, and Join} \\
        \hline 
        Document & \makecell[c]{Filter (with dot processing), Project, \\ Sort, Limit, Aggregate, Union, Join, and Unwind} \\ 
        \hline 
        Array & \makecell[c]{Element-Wise Arithmetic, Matrix Multiplication, \\ Transposition, Aggregation, Window, \\ Sub-Array, Spatial Join, and Array Build} \\ 
        \hline 
        Inter-Model & \makecell[c]{Model Conversion and Join} \\
        \bottomrule
    \end{tabular}
    \label{table:operations}
\end{table}
\section{Query Processing with Integrated Engines}
\label{sec:storage}

\subsection{Query Execution Flow}
\label{subsec:query_proc}

When a user submits a multi-model query through \systemnamens's querying interface, the system processes it through several stages.
First, the query planner interprets the query, performing binding operations to create a logical plan represented as a directed acyclic graph (DAG). 
This logical plan consists of nodes representing operations and edges indicating data flow relationships between producer and consumer operations, capturing all operations across different data models and their relationships.
The plan is then partitioned based on the data models involved. 
Each partition contains operations specific to a particular model (relational, document, or array) that will be executed by the corresponding storage engine.
DuckDB handles both relational and document operations~\footnote{
    We enhanced DuckDB with VelocyPack~\cite{velocypack}, a JSON serialization format, and added specialized JSON processing operations to optimize document processing in DuckDB.}
, while PreVision processes array computations. 
A bridge module sits between the execution engines, handling inter-model conversions and join operations.

\keyword{Partitioning}
Given a logical plan, \systemname constructs a partition DAG comprised of partitions and their dependencies to group model-specific logical operations.
A partition contains nodes with the same data model, and the only edges connecting the nodes.
Inter-partition edges preserve the direction of the underlying operation graph, connecting partitions that produce data to their consumers.

\systemname uses a very simple heuristic approach to partition a logical plan.
The query planner constructs partitions bottom-up; it combines compatible partitions into coarse-grained units.
The partitioning process begins by creating an initial partition DAG where each operation resides in its own partition. 
The planner then performs bottom-up merging, starting from partitions with no producers. 
For each such partition, the planner traverses the partition graph using depth-first search.
During traversal, it evaluates potential merging opportunities between the current partition and its consumers, performing merges when the following conditions allow:
\begin{enumerate}
    \item 
    Both partitions must contain operations of the same model.
    
    \item 
    The merged partition must have exactly one output node that does not have consumer nodes belonging to the partition, thus ensuring query executability.
    
    \item 
    The merging operation must not create cycles in the graph.

\end{enumerate}

Once traversal completes, the resulting merged partitions are sorted topologically to determine the execution order between partitions.
The ordered partitions are now ready for submission to their respective execution engines.

\keyword{Query Submission}
\systemname constructs queries for storage engines from partitions using the querying interface provided by each engine. 
Since PreVision's querying system is modeled as a DAG similar to \systemnamens's approach, \systemname simply transforms its plan to PreVision's query graph before submission. 
DuckDB requires a different approach, as its query shape is tree-based rather than DAG-based. 
To achieve this transformation, \systemname uses a simple materialization approach to decompose a DAG-based partition into a tree structure for proper DuckDB query submission.

When transforming a DAG query into a tree structure for the relational engine, \systemname begins by identifying nodes without producers and traverses the graph bottom-up from these starting points.
During traversal, \systemname searches for nodes that have more than one consumer, as these nodes create the DAG structure.
Upon finding a node with multiple consumers, \systemname detaches the entire subtree rooted at that node and stores it in a tree list maintained throughout query execution. 
\systemname then creates aliased nodes to represent this subtree and connects each aliased node to the original consumers. 
At the end of the traversal, the graph is ultimately transformed into a tree structure containing both regular and aliased nodes.

For query execution, \systemname submits each constructed query tree to the relational engine sequentially. 
First, \systemname processes the trees from the tree list, converting each into a DuckDB query tree. 
Since the roots of these trees originally had multiple consumers, their results are materialized. 
Any nodes that consume these materialized results are configured to reference them appropriately. 
Finally, \systemname submits the main tree to DuckDB, completing the query execution.

\subsection{Unified Buffer Pool}
\label{subsec:buffer}

The storage engines in \systemname are integrated into a unified buffer pool to optimize memory utilization.
In the original versions of DuckDB and PreVision, each storage engine has its own buffer space, requiring dedicated allocations of the host's memory.
Since neither storage engine supports dynamic buffer space adjustment during query runtime, their buffer sizes must be determined before query execution.
This limitation leads to lower buffer pool utilization by each storage engine, resulting in a poor buffer-hit ratio.

The buffer utilization issue becomes particularly problematic when an array model is involved.
Consider the non-negative matrix factorization algorithm producing a significant amount of intermediate matrices~\cite{koo2024prevision}.
The matrices take a large portion of buffer space, leading to buffer pool exhaustion.
It results in high-volume disk spills, which are aggravated when the buffer size is limited
(this trend is observable  in Section~\ref{sec:eval}).
Thus, it is beneficial to allow storage engines to utilize the entire buffer space.

The unified buffer pool of \systemname provides a memory allocator and buffer pool utility functions.
The memory allocator offers allocation functions such as \code{malloc} and \code{free}, and all buffer allocation requests are redirected to this allocator.

The utility functions include two essential buffer management operations: the \code{Add} function registers allocated objects with the buffer pool, and the \code{Evict} function requests the buffer pool to clear cached objects (under the LRU policy) until the requested amount of free memory space becomes available. 
Additionally, the buffer pool requires a storage engine to implement two callback functions: \code{isEvictable} and \code{doEviction}. 
The \code{isEvictable} function determines whether a buffer object can be removed from the buffer pool (\eg, by checking that no readers are currently referencing the object). 
The \code{doEviction} function handles cleanup of the buffer object (\eg, writing array data to disk in the array file format). 
When the \code{Evict} function selects a buffered object for removal, it first calls the \code{isEvictable} function, and if the object is evictable, it then calls the \code{doEviction} function.

The buffer managers of DuckDB and PreVision are modified to attach to the unified buffer pool. 
The buffer managers of the engines invoke the \code{Add} and \code{Evict} functions when a buffer object is created and when insufficient space exists in the buffer pool, respectively.
The engines register their callback functions with the unified buffer pool, enabling it to skip the currently used buffer objects and properly handle cleanup. 

\section{Multi-Stage Hash Join}
\label{sec:join}

In multi-model analytic workloads, data joins between an array model and another model (relation or collection) occur frequently. 
For example, in a benchmark for multi-model analytic tasks~\cite{kim2022m2bench}, four out of six array-involving tasks demonstrate joins between the array and another model. 
These cases utilize \textit{spatial join}, which matches dimension values in the array model with attribute values in the other model.

A naive approach to performing such joins employs relational join operations, which necessitates converting array input data into relational format. 
This conversion incurs significant data copying costs, a problem that worsens as array volumes increase, given their inherent multi-dimensional nature with huge amounts of cells.

Array processing and management systems typically adopt tiled storage architectures to leverage data access locality~\citeN{koo2024prevision}.
Unlike traditional heap files in relational storage engines, which store records unordered, tiled storage organizes cells based on their coordinates, partitioning the array into rectangular tiles that function as pages or blocks.
However, existing relational join operations do not exploit the spatial locality of tiled storage, leading to suboptimal performance with numerous data copies.
Hence, a join method should leverage the array format structure to utilize the spatial locality to improve performance while minimizing data copies.

To improve performance by fully harnessing spatial locality, we propose \textit{multi-stage hash join} (\mmjoin), a binary inter-model equi-join method that combines arrays and other models in their native storage, eliminating costly data transformation.
\mmjoin optimizes array access patterns to minimize disk I/O and, in some cases, produces output without accessing all array data.
The joined result is produced in one of the input models.


\begin{algorithm}[t]
\small
\SetAlgoLined
\SetKwProg{MSHJ}{\texttt{MSHJ}}{}{}
\DontPrintSemicolon

\MSHJ{$\mathit{(Relation: R, Array: A)}$}{

    \nl S = R
    
    \tcp{Building phase}
    \nl \ForEach{ $d \in \{0, \ldots, D-1\}$ }{
        
        \lnl{algo:mshj:bucketbuild} 
        $B^{d} = \{\emptyset_i \mid \text{size}(B^{d}) = \lceil AS_d / TS_d \rceil \}$
        
        \nl \ForEach{ $r \in S$ }{
            \lnl{algo:mshj:bucketinsert1} 
            i = $f_d(v_d \in r)$

            \lnl{algo:mshj:bucketinsert2} 
            $B^{d}_{i}\text{.add}(r)$
        }
    
        \nl $S = B^{d}$
    }
    
    \tcp{Probing phase}

    \nl \ForEach{ $r \in S$ }{

        \lnl{algo:mshj:tc} 
        $TC = (f_0(v_0),\ldots,f_{D-1}(v_{D-1}))$
        

        \lnl{algo:mshj:cc} 
        $CC = (v_0 \bmod TS_0,\ldots, v_{D-1} \bmod TS_{D-1})$

        \lnl{algo:mshj:buf1}  
        \If{$t.TC \ne TC$}{
            \tcp{Get a tile if new visiting tile}
            
            \lnl{algo:mshj:buf2} 
            $t = A.\text{pin}(TC)$
        }

        \tcp{Get a cell and produce a join result}
        \lnl{algo:mshj:emit} 
        emit($r \Join t\text{.get($CC$)}$)
    }
}

\caption{\mmjoin between a relation and an array}
\label{algo:mshj}

\end{algorithm}

Algorithm~\ref{algo:mshj} describes the \mmjoin procedure between array and relational data models, taking data from both models as input.
Let $AS$ denote the array size, $TS$ the tile size, and $D$ the number of dimensions.
For clarity, we use a relation as the non-array input model, but a join between a collection and an array can be performed in the same way, as documents in collections can also contain dimension attributes.
We use 0-based indexing throughout this section.

The initial step of \mmjoin is reordering relation records to align with the tiled structure.
This reordering is necessary because evaluating join conditions between unordered relation records and array cells would require random access to array tiles, resulting in high disk I/O costs.

Similar to traditional hash joins, \mmjoin operates in two phases: building and probing. 
However, it introduces key differences: the building phase consists of $D$ sequential stages (one for each array dimension), and each stage employs a different hash function.

The building phase of \mmjoin processes each dimension sequentially using its corresponding hash function, defined as $f_d(v_d)$ = $\lfloor v_d / TS_d \rfloor$, where $v_d$ is the relation attribute and $TS_d$ is the tile size for dimension $d$.
At each stage, order-preserving buckets $B^{d}$ are initialized, with the number of buckets matching the number of tiles along that dimension (Line~\ref{algo:mshj:bucketbuild}).
The stage then scans all relation records, applying the hash function $f_d$ to compute the target tile coordinate $i$ for $v_d$ (Line~\ref{algo:mshj:bucketinsert1}).
Each record is inserted into the $i$-th bucket of $B^{d}$ (Line~\ref{algo:mshj:bucketinsert2}).
After each stage, the records are stored in buckets while preserving their original order.
These buckets are then sequentially scanned in ascending order of $i$ and fed into the next stage.
This process continues through all dimensions until the final set of buckets corresponds to the number of tiles in the last dimension $D-1$.
At the end of the building phase, records in each final bucket are ordered according to dimensions $D-2, D-3, \ldots, 0$, similar to the radix sort.

In the probing phase, \mmjoin sequentially scans records from each bucket in ascending order of $i$.
For each record, the tile coordinates $TC$ and the cell coordinates $CC$ are calculated (Line~\ref{algo:mshj:tc} - Line~\ref{algo:mshj:cc}).
Using these coordinates, the corresponding tile $t$ and cell $t.get(CC)$ are accessed.
If the tile differs from the previously accessed one, a new tile is pinned (Line~\ref{algo:mshj:buf1} - Line~\ref{algo:mshj:buf2}).
The final result is then produced by concatenating the record with the corresponding cell values (Line~\ref{algo:mshj:emit}).
In PreVision, tiles are stored in one of three formats: dense (cells in contiguous regions), sorted coordinate list (COO), or compressed sparse row (CSR).
For dense tiles, probing uses direct cell access, whereas for sorted COO and CSR, probing is done in logarithmic time using binary search to locate matching cell coordinates.
During the matching step, tiles are cached, and each tile is pinned {\em exactly once}.
This is ensured by the bucket organization, which groups record accesses by tile coordinates.


\begin{figure}[t]
    \centering
    
    \begin{subfigure}{\linewidth}
		\includegraphics[width=0.925\textwidth,keepaspectratio]{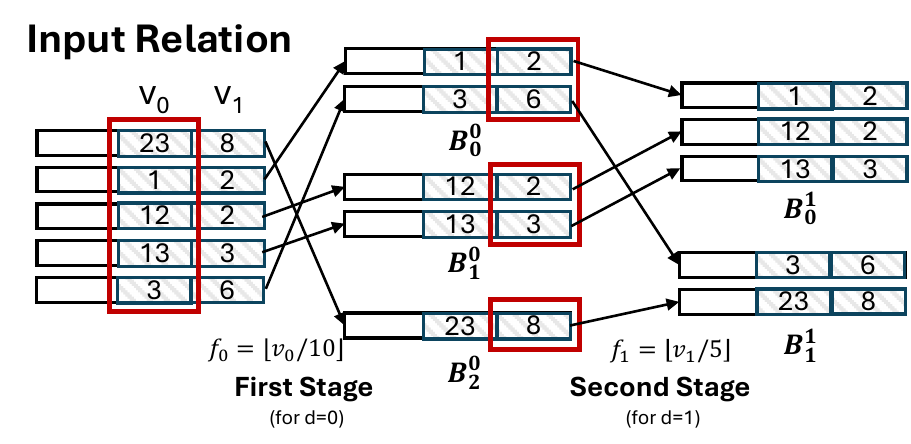}
        \caption{Building phase}
        \label{fig:mshj_ex:1}
	\end{subfigure}

    \begin{subfigure}{0.95\linewidth}
		\includegraphics[width=0.925\textwidth,keepaspectratio]{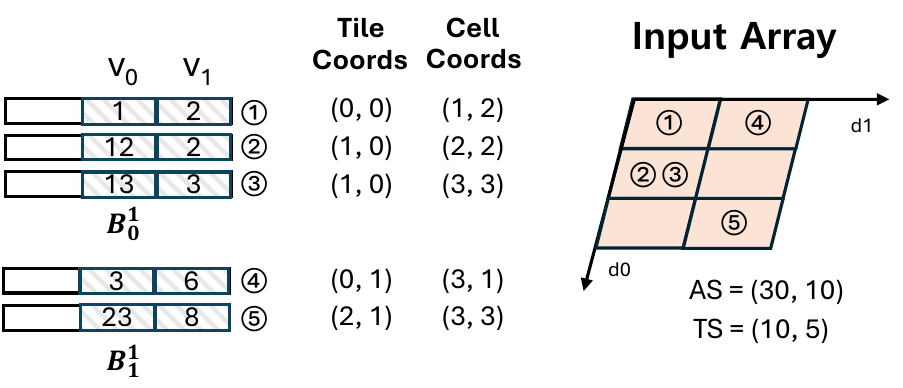}
        \caption{Probing phase}
        \label{fig:mshj_ex:2}
	\end{subfigure}

    \caption{An example of the \mmjoin procedure}
    \Description{An example of the \mmjoin procedure}
    \label{fig:mshj_ex}
\end{figure}

\begin{qedexample}
Figure~\ref{fig:mshj_ex} illustrates a working example of \mmjoin for a two-dimensional spatial join. 
The input relation contains five records, each with $v_0$ and $v_1$ as dimension attributes. 
The input array has dimensions of 30 $\times$ 10 with a tile size of 10 $\times$ 5.

As shown in Figure~\ref{fig:mshj_ex:1}, the building phase begins with bucketing records based on the first dimension ($v_0$). 
Since the array and tile sizes in the first dimension are 30 and 10, respectively, the hash function $f_0$ is defined as $\lfloor v_0/10 \rfloor$, and three buckets are initialized. 
Each record's $v_0$ value is hashed to determine its bucket placement. 
For example, the first record with $v_0 = 23$ is assigned to bucket $B^{0}_{2}$ (calculated as $\lfloor 23/10 \rfloor = 2$). 
After processing all records, the second stage initializes two buckets for the second dimension, and records are inserted while preserving the ordering established in the previous stage.

Once bucketing is complete, the probing phase begins. 
For each record, both tile coordinates and cell coordinates are computed. 
Tile coordinates are calculated as $(\lfloor v_0/10 \rfloor, \lfloor v_1 / 5\rfloor )$, while cell coordinates within the tile are $(v_0 \bmod 10, v_1 \bmod 5)$.
For instance, the last record in Figure~\ref{fig:mshj_ex:2} yields tile coordinates of $(\lfloor 23/10 \rfloor, \lfloor 8/5 \rfloor ) = (2, 1)$ and cell coordinates of $(23 \bmod 10, 8 \bmod 5) = (3, 3)$.
After identifying the corresponding tile, the algorithm retrieves the cell for the record's coordinates. 
If a valid cell exists, it is combined with the record and emitted as part of the join result. 

The circled numbers on the records and tiles represent their access order.
As the figure shows, the algorithm's sequential processing ensures each tile is accessed exactly once, eliminating the need for costly repeated tile buffer operations.
Notably, this example demonstrates how \mmjoin avoids scanning the entire array by accessing only the necessary tiles, minimizing disk I/O while producing the correct output.
\end{qedexample}


With buckets implemented as hash tables, \mmjoin achieves $O(DN)$ time complexity for the building phase, where $D$ denotes the number of dimensions and $N$ represents the number of records in a relation.
The probing phase complexity varies depending on the tile format used on the array side. 
When using dense format, probing occurs in $O(D)$ time, resulting in an overall time complexity of $O(DN + DN) = O(DN)$. 
Conversely, in PreVision, cells in sparse formats are stored in sorted order.
Thus, binary searching on cell coordinates requires $O(Dlog M)$ time, where $M$ represents the number of cells in the array side. 
This yields an overall time complexity of $O(DN + DN log M) = O(DN log M)$.
For disk read, blocks on the table side must be read $D$ times, while tiles on the array side are accessed only once.
Consequently, the algorithm incurs $O(DB + T)$ block accesses to read, where $B$ denotes the number of blocks in the relation and $T$ indicates the number of tiles in the array.
The disk write cost is $O(B)$ if a relational output is requested, while an array output incurs $O(T)$ cost.
Since $D$ is generally small in practical applications, it can be treated as a constant factor.

Since inter-model joins are performed in the bridge component, which can access both DuckDB and PreVision, it can write join output directly to both data layouts.
When the output is an array, cell writing processes for the output array also benefit from an optimized tile access pattern because input and output tiles have equal sizes in PreVision.
In contrast, if inter-model joins producing array output are executed through relational joins, the join results must be converted to array format afterward.
\section{Evaluation}
\label{sec:eval}

\begin{figure*}[ht!]
	\centering
    \begin{subfigure}{0.5\textwidth}
		\includegraphics[width=\linewidth,keepaspectratio]{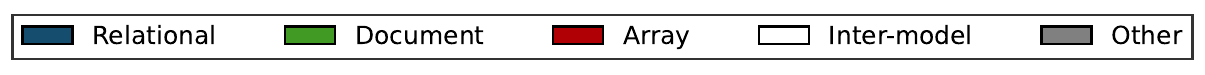}
	\end{subfigure}
    
    \begin{subfigure}{0.16\textwidth}
		\includegraphics[height=3.5cm,keepaspectratio]{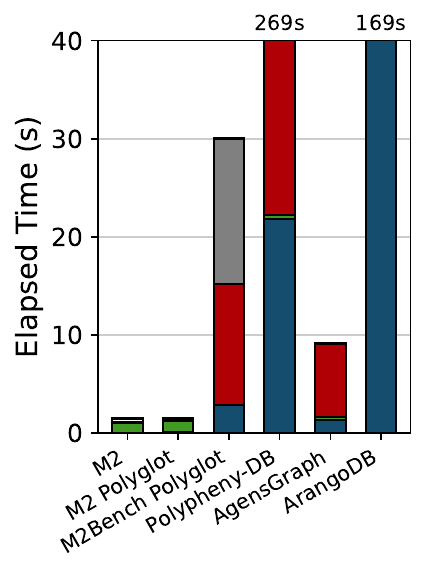}
        \vspace{-0.8em}
		\caption{Task 0}
		\label{fig:eval_main_t0}
	\end{subfigure}
	\begin{subfigure}{0.16\textwidth}
		\includegraphics[height=3.5cm,keepaspectratio]{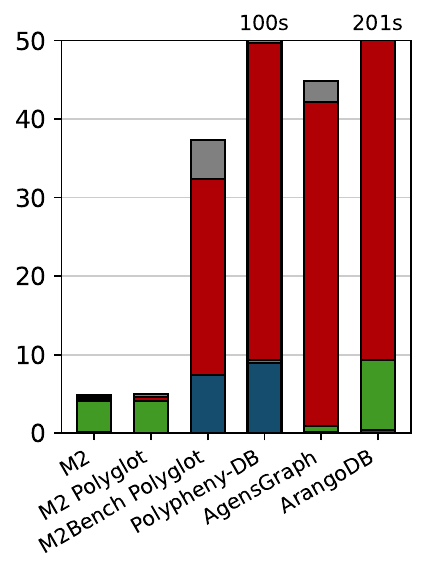}
        \vspace{-0.8em}
		\caption{Task 2}
		\label{fig:eval_main_t2}
	\end{subfigure}    
    \begin{subfigure}{0.16\textwidth}
		\includegraphics[height=3.5cm,keepaspectratio]{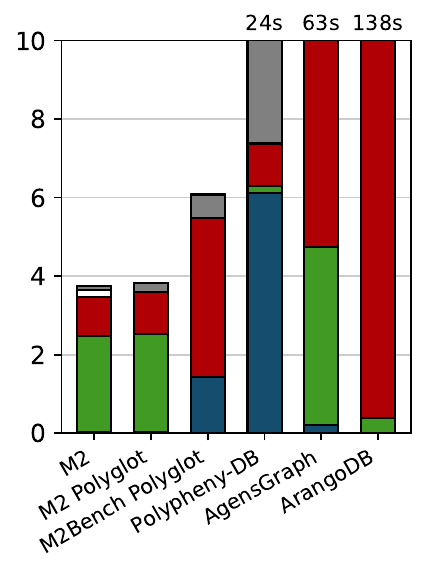}
        \vspace{-0.8em}
		\caption{Task 9}
		\label{fig:eval_main_t9}
	\end{subfigure}
    \begin{subfigure}{0.16\textwidth}
		\includegraphics[height=3.5cm,keepaspectratio]{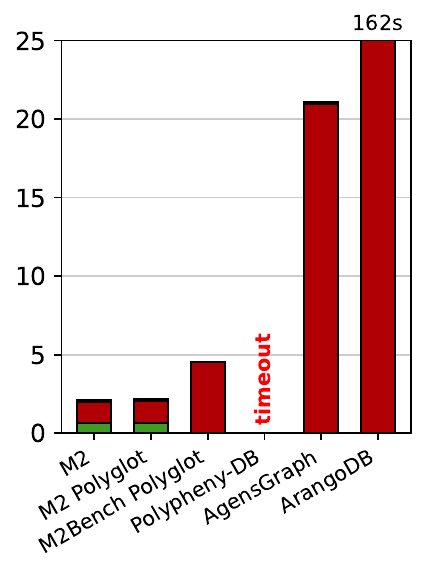}
        \vspace{-0.8em}
		\caption{Task 14}
		\label{fig:eval_main_t14}
	\end{subfigure}
    \begin{subfigure}{0.16\textwidth}
		\includegraphics[height=3.5cm,keepaspectratio]{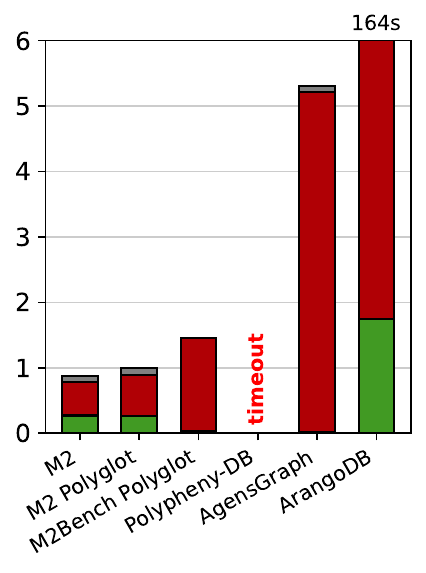}
        \vspace{-0.8em}
		\caption{Task 15}
		\label{fig:eval_main_t15}
	\end{subfigure}
    \begin{subfigure}{0.16\textwidth}
		\includegraphics[height=3.5cm,keepaspectratio]{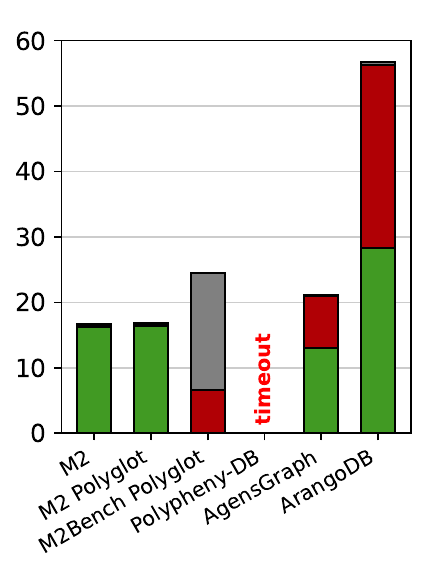}
        \vspace{-0.8em}
		\caption{Task 16}
		\label{fig:eval_main_t16}
	\end{subfigure}

	\vspace{-0.4em}
	\caption{Overall comparison}
    \Description{Overall comparison}
	\label{fig:eval_overall}
\end{figure*}

We conducted an evaluation to assess the effectiveness of \systemnamens. 
All our evaluations were conducted on a desktop machine with an Intel i7-9700K CPU, 32GB RAM, and a 1TB SSD running Linux.
All systems used in evaluations were configured to utilize 24GB of memory for their buffer spaces.
Before each experiment, we reset both the system-specific buffer spaces and the operating system's page cache to eliminate any performance advantages from cached data.
We configured the systems to utilize a single thread to see clearer performance characteristics of each storage engine.
For systems using multiple storage engines, we configured them to avoid interleaving query executions and to use the same amount of buffer space for each data model.

\subsection{Comparison with Other Systems}
\label{subsec:eval:comparison}

We used M2Bench~\cite{kim2022m2bench}, a database benchmark program for multi-model analytic workloads, as its model coverage aligns with \systemnamens's supported models: relational, document-oriented, and array. 
From the complete M2Bench suite, we chose six tasks that require both execution engines of \systemnamens, ensuring a comprehensive evaluation of our multi-model architecture.

\keyword{Comparison Systems}
We selected several systems for comparison with \systemnamens: ArangoDB~\cite{arangodb}, AgensGraph~\cite{agensgraph}, Polypheny-DB~\cite{polypheny}, a polyglot implementation provided with M2Bench~\cite{kim2022m2bench} (M2Bench Polyglot), and another polyglot implementation developed by us (M2 Polyglot).
SurrealDB~\cite{surrealdb} and OrientDB~\cite{orientdb} were omitted due to limited support for join operations, which are crucial for analytical processing.

Polypheny-DB was configured to utilize MonetDB~\cite{idreos2012monetdb} for the relational model and MongoDB~\cite{mongodb} for the document-oriented model.
Since Polypheny-DB did not support a connector for an array database system, MonetDB was used for array data processing.

M2Bench Polyglot was a hard-coded implementation specifically designed for M2Bench tasks. 
It utilized multiple database systems (one for each data model) with a client program.
While model-specific operations were executed in their corresponding database systems, the client program handled inter-model joins and data conversions, serving as the coordination point between different data models.
The M2bench Polyglot was configured to use MySQL~\cite{mysql}, MongoDB, and SciDB~\cite{SciDB} for the relational, document-oriented, and array models, respectively.
To observe the pure performance advantages of the \systemnamens's architecture, we developed M2 Polyglot using the enhanced version of DuckDB and PreVision, each of which handled the relational and document models and the array model, respectively.
Table~\ref{tbl:sys_vers} shows each system's version.

In both polyglot implementations, join operations between distinct storage engines were executed in the following ways.
In array-involving joins, data from the non-array model was iterated, and the corresponding array cell was matched directly.
Otherwise, the data were joined using the nested-loop method.

\begin{table}[t]
    \small
    \centering
    \begin{tabular}{c|cccccccc}
        \toprule
        \textbf{System} & Polypheny-DB & AgensGraph & ArangoDB & MySQL \\
        \hline
        \textbf{Version} & 93fbff6 & 2.14~\footnote[1]{} & 3.12.3 & 9.2.0 \\
        \hline
        \textbf{System} & MongoDB & SciDB & MonetDB & DuckDB \\
        \hline
        \textbf{Version} & 8.0.4 & 19.11.5 & 11.53.9 & 1.0~\footnote[2]{} \\
        \hline
    \end{tabular}
    \vspace{2ex} 
    \caption{System versions}
    \label{tbl:sys_vers}
    \vspace{-5ex} 
\end{table}

\footnotetext[1]{AgensGraph 2.14 is based on PostgreSQL 14.}
\footnotetext[2]{The modified version is forked from this version.}

\keyword{Configuration}
For array tasks with iterative algorithms, we set the number of iterations to one, as it was enough to observe performance characteristics.
The array data were initially chunked into about 5 megabytes tiles for the engines utilizing the tiled storage.
As an exception, for workloads involving dense matrix multiplication operations, the array data for SciDB were split into 1000 $\times$ 1000 tiles because SciDB imposed a size limit of 1024 for these operations.
Since the benchmark includes queries for a property graph data model, which \systemname does not directly support, we handled the data with a relational storage engine.
Graph data was stored in relational edge and node tables.
For tasks requiring simple one-hop pattern matching, we substituted graph operations with equivalent relational selection queries on edge tables.
We excluded tasks requiring more complex graph operations, such as multi-hop pattern matching or shortest path finding. 

\keyword{Evaluation Results}
Figure~\ref{fig:eval_overall} presents the execution times for benchmark tasks across all evaluated systems with scaling factor one. 
Following the profiling rules of M2Bench, we segment execution times by data model, shown as different colors in the stacked bars: blue for relational operations, green for document operations, and red for array operations. 
Inter-model operations executed by the bridge module of \systemname are represented in white.
For the polyglot implementations, coordinator overhead and communication costs between the coordinator and underlying database systems are both categorized as ``other.''
Any remaining execution time not attributable to a specific model's processing is also included in ``other''.
The overflowed bars display the elapsed times above the bars.
We denote a timeout if the execution time exceeds one hour.

\systemname accelerated query execution by up to $188 \times$ compared to all other systems across all benchmark tasks.
Processing times for relational and array data models were consistently lower in \systemname compared to the alternatives, excluding M2 Polyglot, attributed to its use of specialized engines: the DuckDB engine for relational and PreVision for array operations.

\systemname delivered comparable performance to M2 Polyglot with speedups of  $0.97 - 1.15 \times$, attributed to using the same storage engines.
Polypheny-DB and M2Bench Polyglot showed significant communication overhead by coordinator programs.
In the case of Polypheny-DB, the overheads took about 20\%, 50\%, and 70\% of the total elapsed times for Tasks 0, 2, and 9, respectively. 
While Polypheny-DB and M2Bench Polyglot were required to use inter-process communication between storage engines and coordinator programs, M2 Polyglot embedded DuckDB and PreVision in the same memory space, resulting in lower communication costs.
This result supported our main idea: the integration of multiple specialized storage engines can effectively deliver performance while reducing communication overhead.
Nevertheless, M2 Polyglot requires manual coordination between engines using their own interfaces.
\systemname provides an integrated interface with a unified buffer pool and \mmjoinns; their effectiveness is discussed later.

\begin{figure*}[t]
	\centering
    \begin{subfigure}{0.6\textwidth}
		\includegraphics[width=\linewidth,keepaspectratio]{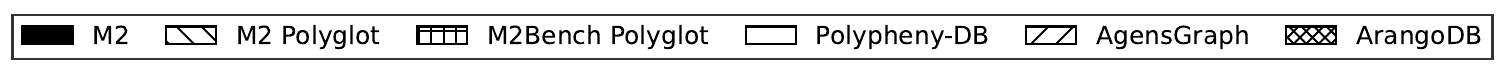}
	\end{subfigure}
    
    \begin{subfigure}{0.30\textwidth}
		\includegraphics[height=3cm,keepaspectratio]{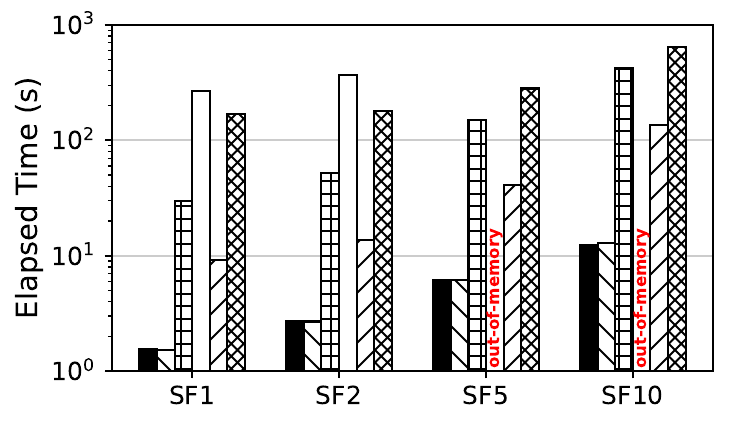}
		\caption{Task 0}
		\label{fig:eval_sf_t0}
	\end{subfigure}
    \begin{subfigure}{0.30\textwidth}
		\includegraphics[height=3cm,keepaspectratio]{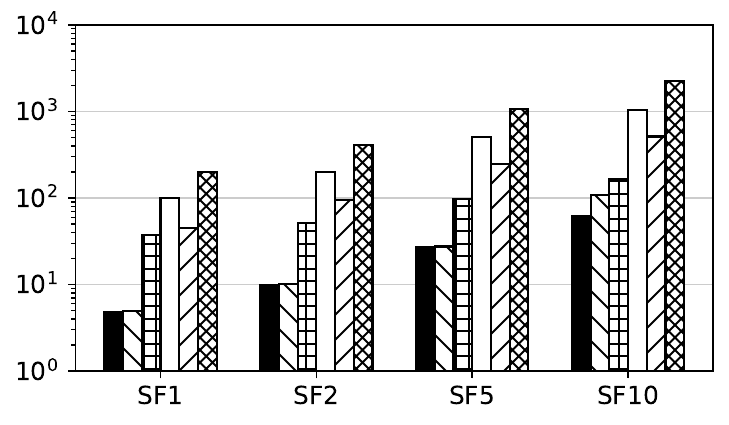}
		\caption{Task 2}
		\label{fig:eval_sf_t2}
	\end{subfigure}
    \begin{subfigure}{0.30\textwidth}
		\includegraphics[height=3cm,keepaspectratio]{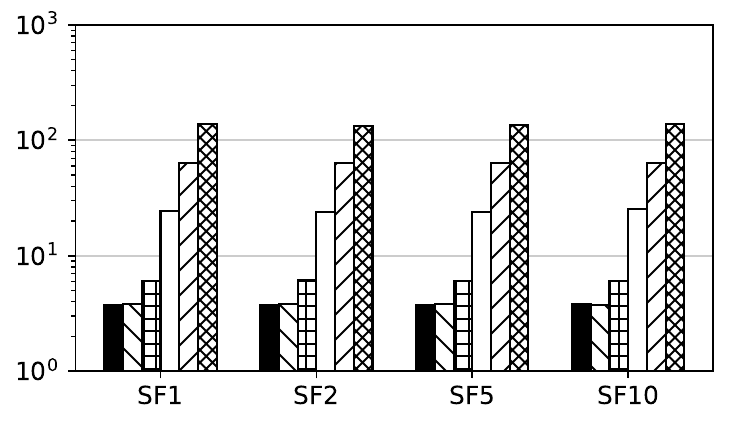}
		\caption{Task 9}
		\label{fig:eval_sf_t9}
	\end{subfigure}
	
    \begin{subfigure}{0.30\textwidth}
		\includegraphics[height=3cm,keepaspectratio]{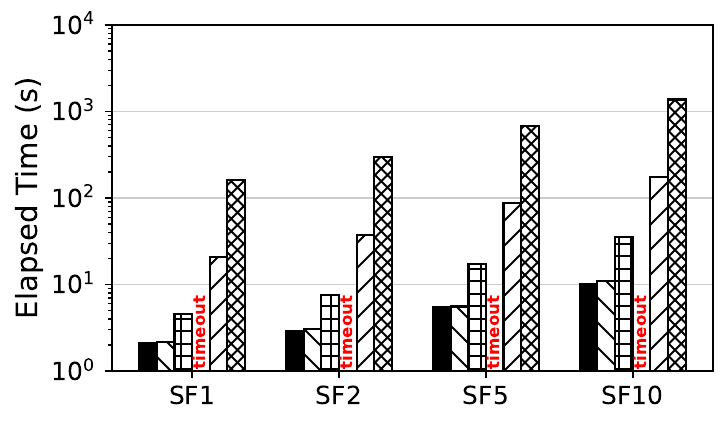}
		\caption{Task 14}
		\label{fig:eval_sf_t14}
	\end{subfigure}
    \begin{subfigure}{0.30\textwidth}
		\includegraphics[height=3cm,keepaspectratio]{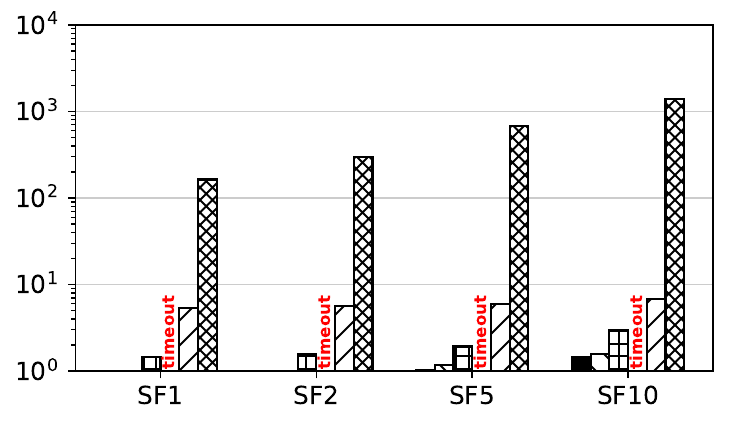}
		\caption{Task 15}
		\label{fig:eval_sf_t15}
	\end{subfigure}
    \begin{subfigure}{0.30\textwidth}
		\includegraphics[height=3cm,keepaspectratio]{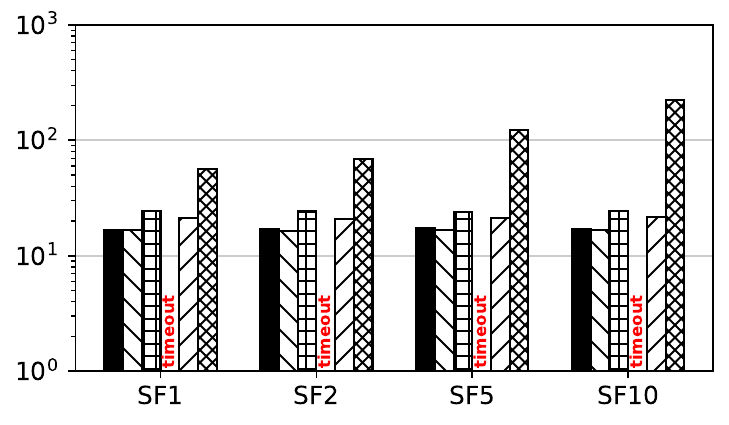}
		\caption{Task 16}
		\label{fig:eval_sf_t16}
	\end{subfigure}
    
	\caption{Performance with varying scale factors}
    \Description{Performance with varying scale factors}
	\label{fig:eval_sf}
\end{figure*}

The M2Bench Polyglot and AgensGraph, which employed a row-oriented storage engine, exhibited inferior performance compared to \systemname in relational processing.
Polypheny-DB presented limited performance for relational processing, despite using a column-oriented relational engine.
During its internal query rewriting step, Polypheny-DB split input queries into numerous sub-queries for storage engine submission, ultimately incurring significant query processing overheads.
ArangoDB demonstrated poor performance when executing joins on relational datasets in Task 0, resulting in considerably longer execution times for relational operations.

For document processing, \systemname did not demonstrate superior performance due to its currently partially optimized implementation.
In Tasks 0, 2, and 9, \systemname performed the VelocyPack serialization excessively, accounting for the majority of the elapsed time.
Tasks 14 and 15 required nearest neighbor searches, but the lack of optimized geospatial indexing led to lengthy document processing times, with geospatial searches consuming 95\% of the document processing time. 
In Task 16, the scanning and filtering of a huge document collection accounted for almost all of the total elapsed time.
As \systemname is a prototype system still under development, we will further optimize the document processing capability of the system.
ArangoDB showed inferior performance in document processing for Tasks 2, 15, and 16. 
In Task 2, the materialization of intermediate results was time-intensive, while in Tasks 15 and 16, index searching operations accounted for a significant portion of the execution time.

M2Bench Polyglot demonstrated inefficient array processing in Tasks 0, 2, 9, and 16.
In Task 16, which involves spatial joins between a document collection and an array, SciDB's performance suffered due to random access of the array data.
In the rest of the tasks, linear algebra operations dominated the array processing requirements, resulting in poor performance by the array engine.
Polypheny-DB exceeded the timeout threshold in Tasks 14, 15, and 16.
Most of each execution time was spent processing window operations which are the bottleneck operations in the tasks.
ArangoDB and AgensGraph stored array data in COO format, necessitating numerous join operations to match cell coordinates during processing.
This led to poor performance for array-intensive tasks in these systems.

\keyword{Varying Scaling Factor}
To evaluate scalability, we conducted assessments using the M2Bench tasks with varying scaling factors.
Figure~\ref{fig:eval_sf} shows the performance results of multi-model solutions across all tasks, with the y-axis representing elapsed times on a logarithmic scale. 
Overall, elapsed times increased as the scaling factor went up, except for Task 9, which remained unaffected by the scaling factor.
Except for M2 Polyglot, \systemname consistently outperformed the other systems at higher scaling factors. 
Notably, \systemname showed superior performance compared to M2 Polyglot at Task 2 with higher scaling factors.
In Task 2, the non-negative matrix factorization algorithm was computed, producing a large volume of intermediate results. 
\systemname was equipped with a unified buffer pool, offering better buffer space utilization compared to M2 Polyglot, where buffer pools for storage engines were physically separated.
This difference resulted in fewer disk spills by \systemnamens, showing better performance.
Polypheny-DB experienced out-of-memory errors in Task 0 with scaling factors over five.
It also incurred timeouts in Tasks 14, 15, and 16 due to query times exceeding one hour.

\subsection{Evaluation for Multi-Stage Hash Join}
\label{subsec:eval:mmjoin}

\begin{figure}[t]
	\centering
    \begin{subfigure}{\columnwidth}
		\includegraphics[width=0.96\columnwidth,keepaspectratio]{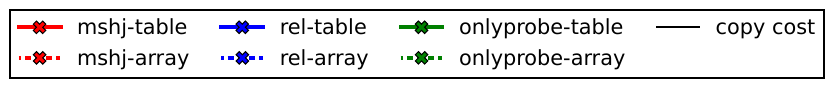}
	\end{subfigure}
    
    \begin{subfigure}{0.49\columnwidth}
		\includegraphics[height=2.8cm,keepaspectratio]{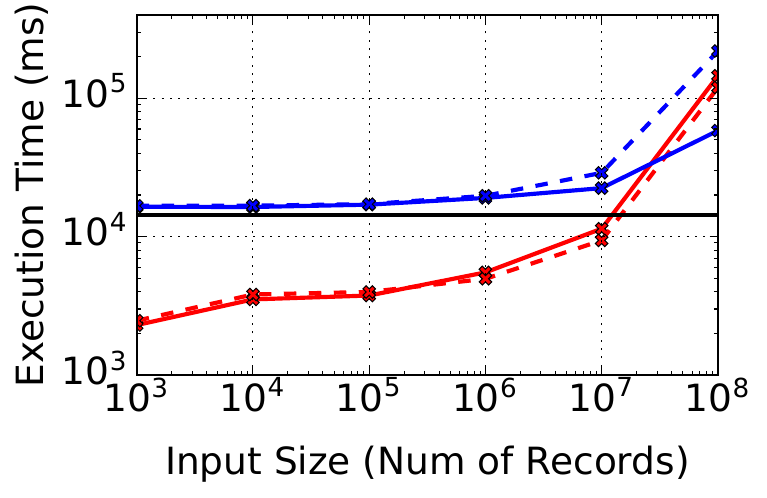}
		\caption{vs. relational joins (3D)}
		\label{fig:eval_mshj:rel}
	\end{subfigure}
    \begin{subfigure}{0.49\columnwidth}
		\includegraphics[height=2.8cm,keepaspectratio]{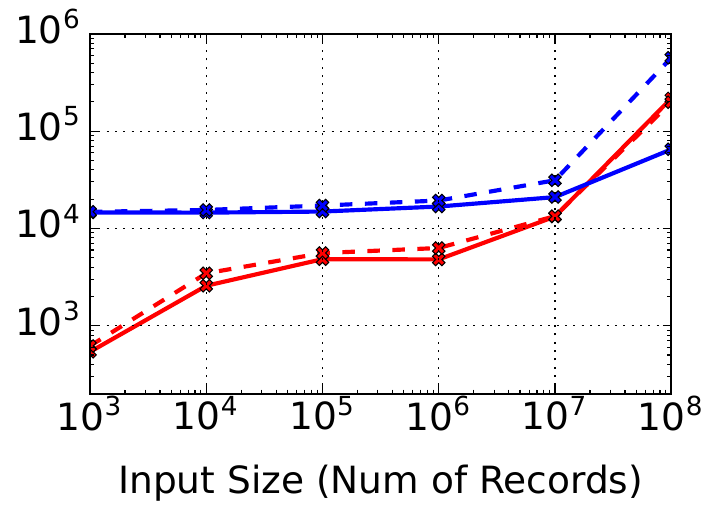}
		\caption{vs. relational joins (4D)}
		\label{fig:eval_mshj:4d}
	\end{subfigure}

    \begin{subfigure}{0.49\columnwidth}
		\includegraphics[height=2.8cm,keepaspectratio]{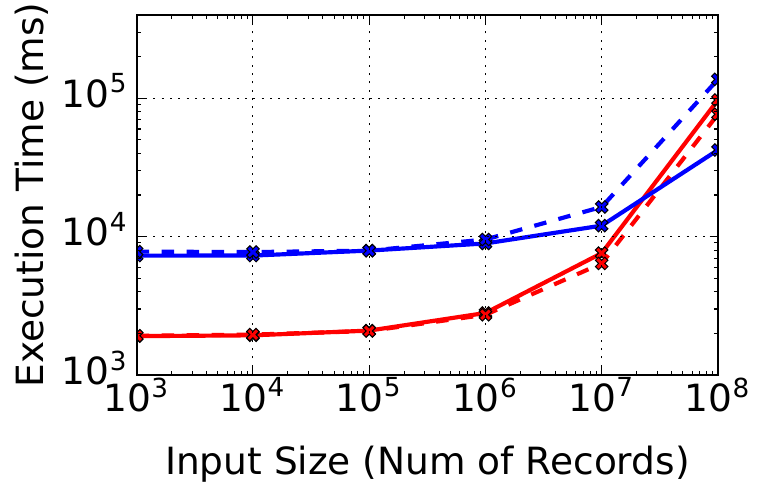}
		\caption{vs. relational joins (2D)}
		\label{fig:eval_mshj:2d}
	\end{subfigure}
    \begin{subfigure}{0.49\columnwidth}
		\includegraphics[height=2.8cm,keepaspectratio]{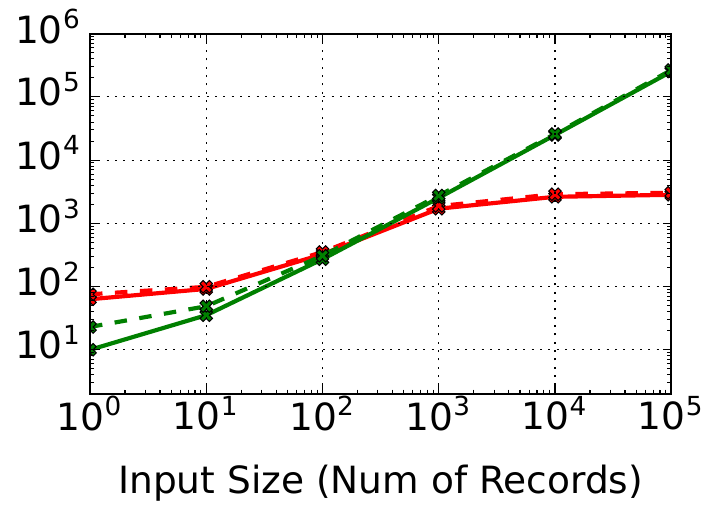}
		\caption{vs. only probing}
		\label{fig:eval_mshj:nobuf}
	\end{subfigure}

	\caption{Performance comparison for \mmjoin}
    \Description{Performance comparison for \mmjoin}
	\label{fig:eval_mshj}
    \vspace{-0.4cm}
\end{figure}

While the overall system evaluation in Section~\ref{subsec:eval:comparison} demonstrates \systemnamens's superior performance, it is difficult to identify the effectiveness of \mmjoin due to the small overhead of inter-model queries in the evaluation.
To directly investigate the effectiveness of \mmjoinns, we conducted a focused comparison between \mmjoin and two alternative approaches for relational-array joins: an indirect method using conventional relational joins and a naive probing-only approach in the bridge component.
For the relational join method, conversion of array data to relational format was necessary before applying standard relational join operations.
If the output model was specified as an array, it transformed the result relation back to the array.
The probing-only approach was implemented to demonstrate the impact of tile access optimization in \mmjoinns.
This method skipped the bucket building phase from \mmjoinns.

All data used in evaluations were synthetically generated.
Arrays were initialized with cells containing a double attribute, with the following specifications: a 2D array of size 10000 $\times$ 10000 with a tile size of 1000 $\times$ 1000, a 3D array of size 500 $\times$ 500 $\times$ 500 with a tile size of 50 $\times$ 50 $\times$ 50, and a 4D array of size 100 $\times$ 100 $\times$ 100 $\times$ 100 with a tile size of 10 $\times$ 10 $\times$ 10 $\times$ 10.
All arrays were initialized in dense format to avoid confusion from cell distribution. 
Relations were prepared with varying record counts ranging from 1 to 100 million. 
Each relation was initialized with dimension attributes matching the domains of the array, plus one double attribute serving as a payload.

Figure~\ref{fig:eval_mshj} illustrates the performance comparison between \mmjoin and the other join methods as the number of records on the relational side varies. 
In the results, ``mshj'', ``rel'', and ``onlyprobe'' prefixes represent the performance of \mmjoinns, relational joins, and our implemented naive approach, respectively. 
The suffixes ``table'' and ``array'' indicate the output model of the join results.

Figure~\ref{fig:eval_mshj:rel} shows that \mmjoin outperformed relational joins when processing fewer than approximately 30 million records, while relational joins with relation output achieved superior performance with larger record volumes. 
This performance crossover occurred because \mmjoin required multiple scanning processes during its bucketing phase. 
As record numbers increased, these building costs grew proportionally, impacting overall performance. 
Meanwhile, the relational join approach incurred significant copy costs when converting arrays to relations (represented by the black line in the figure). 
When array output was required, relational joins faced additional conversion overhead to transform results back to array format, further affecting their performance. 
The output format had a negligible impact on \mmjoin performance. 
The experimental results in Figure~\ref{fig:eval_mshj:4d} and Figure~\ref{fig:eval_mshj:2d} demonstrate trends similar to those in the 3D array experiments, confirming \mmjoin maintains its performance resilience even as the number of processing stages increases.

To evaluate the tile access optimization of \mmjoinns, we compared it with the probing-only version under buffer-exhaustive conditions where only one input array tile could be buffered in memory.
Figure~\ref{fig:eval_mshj:nobuf} presents the performance comparison between these methods using the 3D array.
The probing-only approach performed faster with fewer than about 200 records, but \mmjoin demonstrated superior performance in all other cases.
With small record counts, the random access costs for tiles in the probing-only method did not exceed \mmjoinns's bucketing phase overhead. 
However, as record numbers increased, random access costs became increasingly dominant, significantly degrading the probing-only method's performance.
\mmjoin maintained consistent performance patterns across both buffer-exhaustive and non-buffer-exhaustive environments, highlighting its I/O efficiency regardless of memory constraints.


\section{Related Work}
\label{sec:related_work}

As analytic workloads increasingly span multiple data models, various approaches have emerged to meet these demands.
Multi-model database systems handle multi-model workloads within a single database system~\cite{lu2019multi}.
Polyglot persistence combines different specialized database systems to cope with workloads with multi-model data~\cite{polyglot2022vldb}.
Data lakes~\cite{chess_datalake, terr_datalake} serve as centralized repositories for storing heterogeneous data, usually in open formats like Apache Parquet~\cite{parquet}.
Lakehouse architecture~\cite{Lakehouse} combines the scalability of data lakes with the reliability of data warehouses.

Research has extensively explored in processing multiple data types.
Multicategory~\cite{multicategory} demonstrates a category-theoretic approach to multi-model query processing.
AIDA~\cite{aida} and HADAD~\cite{hadad} optimize hybrid complex queries that combine relational and linear algebra operations.
A line of studies integrates diverse query interfaces using a unified intermediate representation (IR) to optimize query performance.
PyTond transforms NumPy and Pandas queries into an IR and generates SQL from it to efficiently perform the query in the relational database system~\cite{pytond}.
Weld converts library operations into WeldIR and provides a code-generation runtime to execute its IR efficiently~\cite{weld}.
SDQL introduces a funcitonal collection programming language and executor for hybrid relational algebra and linear algebra workloads with semi-right dictionaries~\cite{sdql}.

Fundamental approaches for joining multi-dimensional data include grid files~\cite{nievergelt1984grid} and spatial hash joins~\cite{lo1996spatial}.
However, these methods cannot be directly applied to inter-model joins between relational and array data structures.

Partitioning and optimization for query graphs have been widely studied.
A prominent research direction is acyclic partitioning for graphs, which aims to partition vertices into blocks without cycles among them~\cite{acyclicpartitioning}.
Several works have applied graph partitioning and optimization techniques to queries structured as graphs~\citeN{fang2020optimizing,tensorflow}.
However, none of these approaches are directly applicable to the graph query optimization framework employed in our system.

M2Bench~\cite{kim2022m2bench} is a comprehensive benchmark designed to evaluate multi-model database systems that support relational, document-oriented, array, and property graph data models.
UniBench~\cite{unibench2} focuses on e-commerce workloads and supports relational, key-value, graph, XML, and JSON formats.
BigBench~\cite{bigbench2} initially supported relational and document models for big data analysis and was later extended to include the key-value model.
In this paper, we used M2Bench because UniBench and BigBench do not support the array model.

\section{Conclusion and Future Work}
\label{sec:conclusion}

This paper presents \systemnamens, a multi-model analytic system with specialized storage engines for relational, document-oriented, and array data models. 
The \systemname query plan is partitioned by data models and processed in storage engines optimized for each specific model.
\systemname features two key techniques: {\em multi-stage hash join}, which handles inter-model joins commonly found in multi-model workloads, and {\em unified buffer pool}, which integrates buffer pools of the storage engines to utilize more buffer space.

Developing \systemname represents the first step toward a standalone multi-model database system that treats all data models as first-class entities and processes them using optimized storage engines.
Throughout the development, we explored fundamental approaches to query execution and operation optimization for multi-model query processing.
Several opportunities remain for future work, including sophisticated query optimization techniques, multi-model query languages, and further optimization of inter-model join and conversion operations.

\bibliographystyle{ACM-Reference-Format}
\bibliography{reference.bib}

\end{document}